\documentclass{pnastwo}

\usepackage[dvips]{graphicx}

\usepackage{pnastwof}

\usepackage{amssymb,amsfonts,amsmath}

\newcommand{\partt}[2]{\frac{\partial {#1}}{\partial {#2}}}
\newcommand{\mbar}{\bar{\mu}}

\contributor{Submitted to Proceedings of the National Academy of Sciences of
the United States of America}
\url{}
\copyrightyear{}
\issuedate{\today}
\volume{}
\issuenumber{}

\begin{document}



\title{Spatiotemporal Structures in Aging and Rejuvenating Glasses}
\date{\today}



\author{Peter G. Wolynes
\affil{1}{Department of Physics and Department of Chemistry and Biochemistry,
Center for Theoretical Biological Physics, University of California, San Diego,
La Jolla, CA 92093}}


\contributor{Submitted to Proceedings of the National Academy of Sciences
of the United States of America}

\maketitle

\begin{article}

\begin{abstract} 
  
Complex spatiotemporal structures develop during the process of aging glasses
after cooling and of rejuvenating glasses upon heating. The key to
understanding these structures is the interplay between the activated
reconfiguration events which generate mobility and the transport of mobility.
These effects are both accounted for by combining the random first order
transition theory of activated events with mode coupling theory in an
inhomogeneous setting. The predicted modifications by mobility transport of the
time course of the aging regime are modest. In contrast, the rejuvenation
process is strongly affected through the propagation of fronts of enhanced
mobility originating from the initial reconfiguration events. The structures in
a rejuvenating glass resemble flames. An analysis along the lines of combustion
theory provides an estimate of the front propagation speed.  Heterogeneous
rejuvenation naturally should occur for glasses with free surfaces. The analogy
with combustion also provides a new way of looking at the uptake of diluents by
glasses described by case II and super case II diffusion.

\end{abstract}






\section{Introduction}

It is an excellent, but still imperfect, approximation to think of a structural
glass as being a frozen snapshot of the liquid state. The macroscopic
properties of a glass depend on its preparation history in contrast to those of
an equilibrated liquid. Likewise, the structure of a glass is history dependent
and continues to change even as it is being studied. The
picture of the dynamics of an equilibrium supercooled liquid that emerges from
the random first order transition (RFOT) theory of the glass
transition\cite{lubchenko.2007} can be described, with some poetic license, as
that of a fluctuating mosaic of structures that locally are chosen from a
diverse set of minimum energy patterns which reconfigure through activated
events. In the nonequilibrium aging glass, this mosaic does not just fluctuate
but continues to evolve, eventually to the dynamic mosaic of an equilibrated
liquid sample at a lower temperature\cite{lubchenko.2004}. An aging glass can
also be heated up, ``rejuvenating'' the sample, again eventually reaching a new
dynamic equilibrium\footnote{The term "rejuvenation" is used here as it is in the spin glass community.
The polymer community uses the term to describe the stress induced reversal
of aging. This latter process is distinct from but may be related to the
present heating protocol.}. In this paper, the spatiotemporal structure of the glassy
mosaic in the intermediate states of development of aging or rejuvenating is
explored within the framework of the RFOT theory.

A spatially ultra-local description of an aging glass has already been provided
by Lubchenko and Wolynes\cite{lubchenko.2004}. As in equilibrium, activated
transitions of small regions in the nonequilibrium glass are driven by the
extensive configurational entropy of local structural patterns. Such
transitions (sometimes called ``entropic droplets'') are retarded by the
stability of the local pattern, which varies spatially, giving a distribution
of relaxation times. The range of  existing local stabilization energies in a
small region may be described by a Lagrange multiplier, the fictive temperature
which can be thought of as fluctuating spatially\cite{chamon.2004}. The least
stable local regions typically re-configure first. Upon cooling, the mosaic,
therefore, begins to contain patches of greater stability than it had before.
In this way, the aging glass not only on the average becomes more stable, but
also, for a time, is even more inhomogeneous in its energy distribution than an
equilibrated sample since it was the fast patches that became slow first. The
average behavior leads to non-linear relaxation much like that used in the
Nayaranaswamy-Moynihan-Tool phenomenology which is based on a uniform fictive
temperature\cite{tool.1946,narayanaswamy.1971,moynihan.1976}. The results of
the microscopic RFOT theory for the nonlinearity parameter in the NMT approach
are in good agreement with experiment. In addition, however the RFOT theory
also predicts a qualitative difference from NMT phenomenology. RFOT theory
suggests the patchier mosaic found at intermediate times should lead to
additional ``ultra-slow'' relaxations. These anomalous relaxation processes
have been observed in some experiments \cite{miller.1997}.

The local picture of aging does not completely account for the dynamical
coupling between re-configuring regions: the reconfiguration of a given region
changes the pinning forces acting on its neighboring regions, allowing their
reconfiguration rates, in turn, to change. This effect, may be termed
facilitation, as in the popular kinetically constrained models of
glasses\cite{fredrickson.1984,ritort.2003}, where it is typically the only
effect being modeled. Facilitation was noted as an aspect of aging by Lubchenko
and Wolynes but was not completely analyzed by them. Bhattacharya et al. (BBW)
have shown that the facilitation of activated processes in equilibrium liquids
can be captured mathematically by combining mode coupling theory (MCT) with the
entropic droplets that describe activated transitions in RFOT
theory\cite{bhattacharyya.2005,bhattacharyya.2008}. We exploit this insight to
describe the nonequilibrium spatiotemporal aspects of glasses by using the
combination of MCT with the RFOT mosaic to motivate a continuum theory for the
spatiotemporal coupling between activated events in a glass or supercooled
liquid. An equation for what may be called a mobility field\cite{merolle.2005},
follows from inhomogeneous mode coupling theory within this BBW-based
framework. Unlike existing treatments of inhomogeneous mode coupling theory,
the equation has a spatially varying source term. This source arises from the
activated events, whose dynamics within RFOT depend on the local energy or
fictive temperature. The local energy, in turn relaxes to local-equilibrium at
a rate that depends on the local mobility at the same location. The resulting
coupled equations for a mobility field and for a local energy or fictive
temperature resemble the nonlinear diffusion equations encountered in the
theory of combustion\cite{zeldovich.1944,merzhanov.1988}. In combustion,
chemical kinetics depends strongly on temperature which is increased through
heating from the reaction events themselves, but that
also is transported by conduction. For the aging glass, the coupled
nonlinearities will be shown to have a quantitative but not qualitative effect
on the spatiotemporal evolution of the mosaic. On the other hand, the situation
is quite different for the rejuvenating glass. The rejuvenating glass, like a
highly combustible mixture is unstable and once reconfiguration events are
nucleated they can propagate through a sequence of additional reconfiguration
events, as do exothermic reaction events in a flame.  These flame structures in
the glass speed up the rejuvenation process enormously. While specific
quantitative evidence for such structures has not yet come to light in a
homogeneously rejuvenating glass, there are hints from imaging methods like
those pioneered by Israeloff\cite{israeloff.2007}.  Ediger has observed front
propagation that initiates at the surface of vapor deposited ultrastable glass
upon heating it\cite{swallen.2008,swallen.2008a}.  The speed of the observed
heterogeneous rejuvenation front seems to be consistent with the predictions of
the present theory.

The plan of the paper is as follows. A route to the equations for the mobility
field driven by activated events is described. An estimate for the effect
of facilitation in the ordinary aging situation is obtained. Front
propagation in the rejuvenating glass is then treated using some of the
simpler approximations from combustion theory\cite{merzhanov.1988}.  This
simple treatment allows an estimate of the rejuvenation time for a homogeneous
sample without surfaces.  The origin of heterogeneous rejuvenation within RFOT
is discussed.  Finally the coupling of these spatiotemporal processes to
plasticizer diffusion suggests an explanation within the
RFOT theory framework for the mysterious accelerating penetration of diluents
into polymers known as Supercase II diffusion\cite{frisch.1980}.

\section{Mobility Transport and Mobility Generation in Glasses}

While mode coupling theories are often formulated in a momentum space
representation appropriate for a uniform system, the physics underlying the
equations is as valid for inhomogeneous systems as it is for homogeneous ones.
Long ago the nonlinear coupling of hydrodynamic modes in a fluctuating simple
fluid with surfaces was shown to modify the hydrodynamic boundary
conditions\cite{wolynes.1976}. Coupling of structural density fluctuation modes
in an inhomogeneous situation also has been shown to describe a growing length
scale as the transition to nonergodic behavior is
approached\cite{biroli.2004,biroli.2006}. In structural glass dynamics, mode
coupling theory relates the memory kernel for local structural relaxation to
the behavior of density-density correlation functions. These correlation
functions  in turn depend on the same memory kernel but, in an inhomogeneous
system at locations displaced in space and time. The memory kernel is frequency
dependent, for time translation invariant systems. This frequency dependence is
interesting and complicated, leading to behavior resembling $\beta$ relaxation.
In the present paper we will, however, simplify the analysis by treating the
memory kernel for the explicitly time dependent aging or rejuvenating systems,
as a time and space varying, but locally frequency independent, rate,
$\mu(r,t)$. This rate can be called a mobility field. Biroli et al.,
generalized ideal mode coupling theory to the inhomogeneous situation by
expanding the self-consistent equations in the degree of inhomogeneity and in
gradients to obtain a continuum description\cite{biroli.2006}. Carrying out
this procedure for the memory kernel of mode coupling equation containing an
activated event term, as proposed by BBW, yields an equation of the form

\begin{equation}
  \partt{\mu}{t}  = \bar{\mu} \xi^2 \nabla^2 \mu + c (\nabla \mu)^2 \bar{\mu}
  - \bar{\mu} (\mu-\bar{\mu})
  \label{eqn.1}
\end{equation}

\noindent Here $\mbar$  is the uniform solution of the homogeneous MCT
equations with activated transitions. In the strongly glassy regime MCT effects
are modest.  Therefore $\mbar$ will depend on the local temperature and fictive
temperature, in a way very close to the existing Lubchenko-Wolynes (LW) theory
for uniform systems that accounts for activated transitions alone.

By taking the static limit we can see that the coefficient of the linear
Laplacian term contains a length $\xi$  which is the correlation length of the
4-time correlation function consistent with the Biroli et al. analysis. The
gradient squared term arises because of the nonlinear relation of MCT closure
between memory kernel and correlation functions. The coefficients $\xi^2$ and
$c$, depend on the details of the microscopic mode coupling closures employed.
For simplicity we will choose $c$ so that the locally linearized equation can
be written as a mobility flow equation with a source 

\begin{equation}
  \begin{split}
    \partt{\mu}{t}  =& - \nabla \cdot \mathbf{j}_{\mu} - \bar{\mu} (\mu-\bar{\mu})\\
    \mathbf{j}_{\mu} =& - \bar{\mu} \xi^2 \nabla \mu
  \end{split}
\end{equation}

\noindent In the nonequilibrium glass we must also take the local stabilization as a
variable--equivalently we may say there is a spatially varying fictive
temperature that satisfies an ultra-local relaxation law without conduction
when the finite spatial structure of the activated events is neglected: 

\begin{equation}
    \partt{T_F}{t} = -\mu (T_F - T)  
\end{equation}

\noindent The actual temperature equilibrates by vibrational thermal conduction
rapidly and will be treated as uniform.  In the absence of mobility transport
through a flux term, these equations are equivalent to the Lubchenko-Wolynes
version of the NMT formalism.

To describe the fluctuating fictive temperatures and energies known to occur in
the equilibrium liquid these equations for the mean behavior must be
supplemented by random force terms that are local. Owing to these random forces
the mosaic will fluctuate before a quench or heating process is initiated, then
evolve from these initial fluctuations and then proceed to fluctuate again. The
intensity of the fluctuations in the mobility transport equation reflects the
shot noise of creating entropic droplets while the fluctuating forces in the
fictive temperature equations ensure there is a trend to proper fluctuating
thermal equilibrium. For the situations considered here knowledge of the
intensities, but not of the precise statistics, of these fluctuations is
sufficient.

While $\bar{\mu}$ in pure mode coupling theory undergoes a divergence at the
nonergodicity transition, due to the activated events that cut off the
transition in RFOT theory, $\mbar$  instead becomes a strong, super Arrhenius
function of the fictive temperature. The strong dependence of $\mbar$  on
fictive temperature is the key to the analogy to combustion for rejuvenating
glasses where reaction rates also depend exponentially on the local
temperature. The Lubchenko-Wolynes RFOT analysis of activated events in a
nonequilibrium system suggests that the traditional NMT formulation of $\mbar$
using locally two distinct Arrhenius laws for fictive and ambient
temperature\cite{lubchenko.2004} should be reasonably adequate if the quenches
are not too extreme:

\begin{equation}
  \bar{\mu} (T_F, T) = \mu_0 \exp \left\{ \frac{-x E^{\ddagger}}{T} -
  \frac{(1-x) E^{\ddagger}}{T_F} \right\}
\end{equation}

\noindent When $T=T_F$ this reduces to the Adam Gibbs form, according to RFOT
theory.  The more complete LW RFOT expression containing the configurational
entropy explicitly could equally well be used. In RFOT owing to the
configurational entropy dependence the nonlinearity parameter $x$ is predicted
to be a function of the configurational heat capacity $\Delta C_P$. This result
of RFOT theory agrees well with experiment\cite{lubchenko.2004}. The BBW-MCT
for a uniform system should give only mild modifications to the local LW-RFOT
relation due to the average effects of facilitation while the spatiotemporal
structures considered here from the inhomogeneous version of the theory can
have qualitative effects on the global process of equilibration.

\section{Aging Glasses}

\begin{figure}
  \begin{center}
    \includegraphics[width=0.48\textwidth]{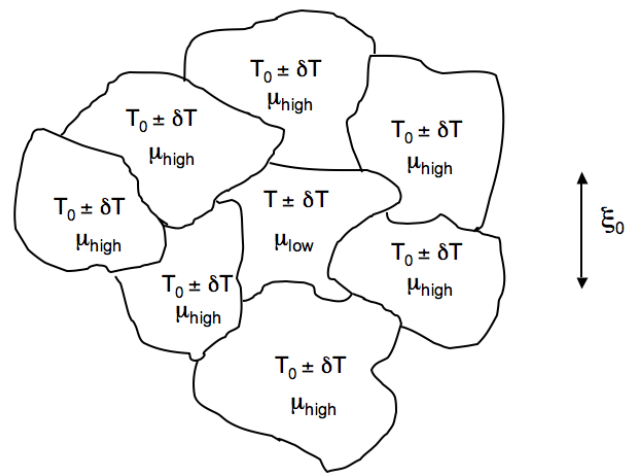}
  \end{center}
  \caption{The mosaic pattern of mobilities soon after a quench from the
  temperature $T_0$ to a lower temperature $T$, as the sample begins to age. A
  few regions of size $\xi_0$ have equilibrated to low fictive temperature
  structures with $T_F$ to near to $T$, while others maintain their higher
  initial fictive temperature. The initially equilibrated regions now have a
  low mobility $\mu_{\mathrm{low}}$, contributing to the "ultraslow"
  relaxations. Mobility transport to these regions also slows approach to
  equilibrium of the neighboring cells but the effect is modest.}
  \label{fig1}
\end{figure}

How is the spatiotemporal pattern in the aging glass after rapid quenching
modified by the mobility transport that was not studied explicitly by LW?
Consider a rapid quench to an ambient temperature $T$ from an equilibrated
sample at a temperature $T_0$ which is then at the average initial fictive
temperature $T_F^{\mathrm{in}} = T_0$ but that also has local $T_F$
fluctuations given by the configurational heat capacity. The initial dynamics
does not involve mobility transport but only activated transitions as already
described by LW.  This initial step generates droplets of radius $\xi_0$, that
are nearly equilibrated. The radius $\xi_0$ is predicted by LW to typically be
about 5 particle spacings, $r_0$, near the laboratory glass transition, $T_g$.
These initial transitions occur at random, starting in those regions where
$T_F$ had fluctuated to a higher than average value. After a reconfiguration
event, each of these regions has come to equilibrium at a fictive temperature
near to the ambient temperature $T$. As shown in figure \ref{fig1} the mosaic,
after some initial transitions has occurred, is more inhomogeneous than a
sample characterized by a fictive temperature with only Gaussian fluctuations
would be. The number of the  reconfigured regions per unit volume during this
initiation phase is thus $(r_0/\xi_0)^{3} (1-\phi(t,T_F ,T))$ where $\phi$ is
the relaxation function of the glass.  Neglecting the possible role of an early
$\beta$  process, the relaxation function for either a supercooled liquid or
glass within RFOT theory is tolerably well approximated as a stretched
exponential $e^{-(\bar{\mu}_0 t)^{\beta}}$ where $\bar{\mu}_0 = \mbar (T_F =
T_0, T)$

Each of the regions that have reconfigured now becomes ultra-slow, since each
one typically has a greater stability than it had before. This is the origin of
the anomalous relaxation component found by Miller and
MacPhail\cite{miller.1997}.  Mobility transport now will affect the
surroundings of each of the initially reconfigured regions. These surrounding
regions relatively speaking, reconfigure more rapidly than the newly
equilibrated regions, being at the higher initial fictive temperature.  The
initial reconfigured regions can thus be considered a static influence on the
relaxation of their surroundings.  Nevertheless if $T$ is considerably lower
than $T_0$, then the mobility of the remaining regions to be equilibrated is
not just $\mbar (T_F,T)$ but is additionally slowed by mobility transport to
the ultraslow inclusions in the mosaic which have mobility
$\mu_{\mathrm{low}}$. As shown in figure \ref{fig1} each of the low fictive
temperature regions affects a region around it, lowering the mobility following
the quasi-static law\cite{stevenson.2008b}

\begin{equation}
  - \xi^2 \nabla^2 \mu + (\mu - \bar{\mu} ) = 0
\end{equation}

\noindent where $\mbar$ is very small within a previously reconfigured region
but large elsewhere.  $\mu$ is decreased significantly only within a length
$\xi$ of a drop. It follows that even in the perturbed surroundings of an
ultraslow cell the rate typically is changed through mobility transport by a
modest factor of roughly two until the reconfigured regions are very close to
merging.  The regions of influence of the initially reconfigured mosaic cells
will overlap when

\begin{equation}
  (2 \xi)^3 \xi_0^{-3} (1-\phi) \approx 1 \Rightarrow \phi \approx  7/8
\end{equation}

\noindent The stretched exponential relaxation implies

\begin{equation}
  t_{\mathrm{\mathrm{overlap}}} \approx \frac{1}{8} \mu_0^{-1}
\end{equation}

\section{Rejuvenating Glasses}

\begin{figure}
  \begin{center}
    \includegraphics[width=0.48\textwidth]{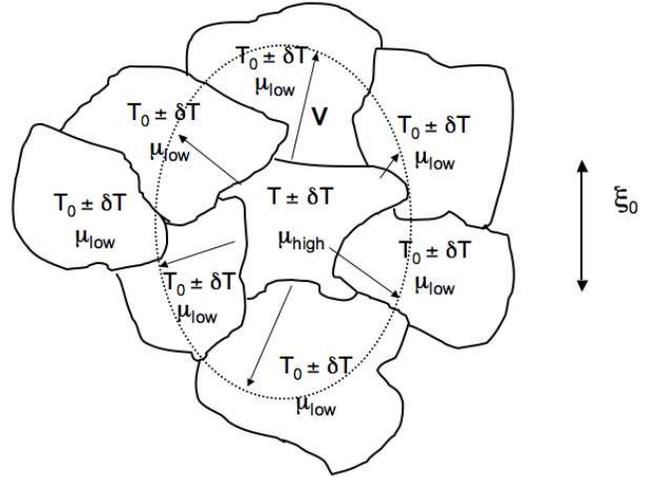}
  \end{center}
  \caption{The mosaic pattern of mobilities for a sample originally
  equilibrated at the temperature $T_0$ which has just been raised to the
  temperature $T$, beginning its rejuvenation. The initially nucleated regions of
  high mobility catalyze the rearrangement of their low mobility neighbors
  leading to a radially propagating front moving at a velocity $v$.}
  \label{fig2}
\end{figure}

The initial stage of rejuvenation of a glass proceeds upon heating very much as
does aging after a quench but with two important differences. Since the ambient
temperature is increased, rather than decreased, the first reconfiguration
events occur faster than they would in the original equilibrated sample (not
slower as in aging). These events still do not happen quite as fast as they
would occur in a sample already equilibrated to the higher ambient temperature
since the sample is initially more stable than an equilibrated high temperature
sample would be. Since the reconfigured regions are now at high fictive
temperature internally their motions become faster. Indeed their mobilities are
exponentially larger (not smaller!) than the mobility of their surroundings as shown in figure \ref{fig2}.
Mobility transport now allows the immediately surrounding material around an
initially reconfigured region to begin to change more rapidly than it would on
its own.  The mobility increase in neighboring regions is autocatalytic and
therefore a front of higher fictive temperature should emanate radially from
each initially rejuvenated center. These growing zones of influence will
quickly overlap and the glass, as a whole will rapidly be equilibrated. This
situation is shown in figure \ref{fig2}. Of course, this simple scenario is
most appropriate when the heating jump is large, which is the case we will
explicitly analyze. The time when overlap of the rejuvenating regions is
reached will be $t_R$, termed the rejuvenation time.

Let us assume the front can propagate stably. The front moving at a velocity
$v$ sweeps out at a volume $(vt)^3$ in a time $t$. Overlap will be reached when

\begin{equation}
  1 \approx (v t_R)^3 \xi_0^{-3} (1 - \phi(t_R, T_F^{\mathrm{in}}, T) )
\end{equation}

\noindent Again approximating $\phi$ by a stretched exponential we obtain

\begin{equation}
  1 \approx (v t_R)^3 \xi_0^{-3} \left( \frac{t_R }{\tau(T_F^{\mathrm{in}}, T) } \right)^{\beta}
\end{equation}

\noindent This implies $t_R$ is essentially a weighted geometric mean of the
time it would take the front to cross a nucleated drop and the original
relaxation time for a system starting at the initial fictive temperature $T_F$
but ambient temperature $T$:

\begin{equation}
  t_R = \left( \frac{\xi_0}{v} \right)^{3/(3+\beta)} \left\{ \tau (T_F^{\mathrm{in}}, T)
  \right\}^{\beta / (3+\beta)}
\end{equation}

\noindent The analogy with combustion allows an estimate of the front velocity.

\section{Rejuvenation Front Propagation and the Combustion Analogy}

\begin{figure*}
  \begin{center}
    \includegraphics[width=0.68\textwidth]{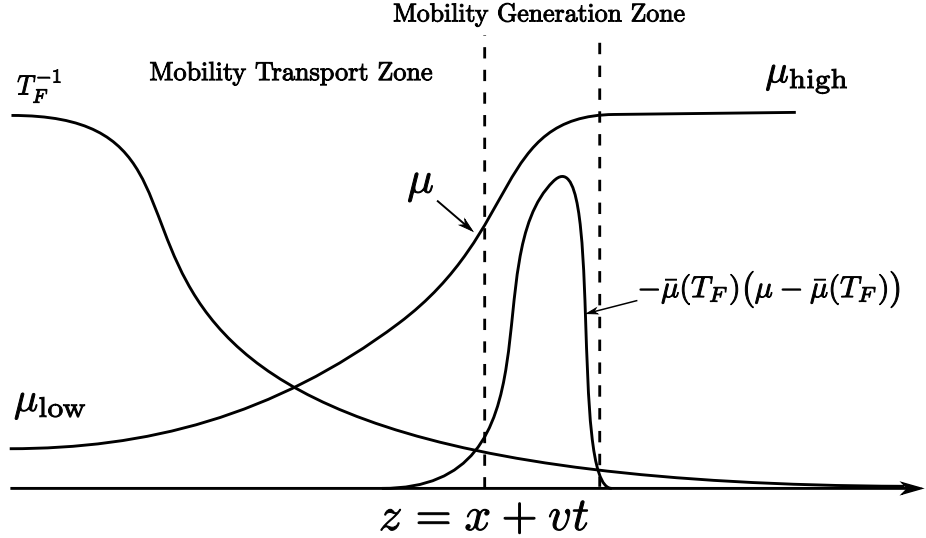}
  \end{center}
  \caption{A sketch of the mobility front in a comoving frame. The mobility
  rapidly rises from $\mu_{\mathrm{low}}$, before the region has equilibrated,
  to $\mu_{\mathrm{high}} \approx \mbar (T)$  after a region has rejuvenated.
  The inverse fictive temperature $T_F^{-1}$ falls while $\mu$ rises.  This
  figure parallels the structure of a flame front, where, however, the
  temperature in the flame is the analog of mobility here and fuel
  concentration is the analog of $T_F^{-1}$. Mobility generation occurs where
  there is both a significant disequilibrium in mobility while the mobility is
  still rather high.}
  \label{fig.3}
\end{figure*}

Consider the coupled mobility/fictive temperature transport equations for a
planar front in which $\mu$ and $T_F$ vary only along a single dimension, $x$.
This situation is shown in figure \ref{fig.3}.  As in the theory of a flame
front\cite{merzhanov.1988}, the coupled equations can be written in a frame
comoving at velocity $v$ in terms of one variable $z$

\begin{equation}
  \begin{split}
  \partt{ }{z} \left( \xi^2 \bar{\mu} \partt{\mu}{z} \right) &- v \partt{\mu}{z}
  - \bar{\mu} (\mu - \bar{\mu}) = 0 \\
  -v \partt{T_F}{z} &- \mu (T_F - T) = 0
  \end{split}
  \label{eqn.12}
\end{equation}

\noindent where $\mbar = \mbar (T_F, T)$.  In the original frame of reference
both $T_F$ and $\mu$ are functions of $z = x + vt$.  $\mu$ and $T_F$ are
graphed schematically in figure \ref{fig.3}. We consider the case where the
mobility at the rejuvenation temperature is much larger than the original
fictive temperature. According to these equations mobility is generated to a
significant degree only in a thin generation zone localized at the front near
$z = 0$. Outside this zone either the fictive temperature has already had time
to equilibrate so $\mu = \mbar = \mu_{\mathrm{high}}$ or else $\mbar$ itself is
small because the sample is too slow to have equilibrated at all yet so the
mobility generation rate is low. The latter condition (no equilibration in the
unperturbed region) is familiar in combustion theory where, in simpler
treatments, it is usually assumed that the unreacted mixture is completely
stable. This assertion is not strictly true since the mixture would eventually
react but on an exponentially long time scale in the absence of autocatalytic
heating by a pre-existing flame. In the current context this infamous ``cold
boundary difficulty'' is not a problem under the condition that rejuvenation
jump is sufficiently large. Flame propagation and front propagation for a
random first order transition correspond to a problem of ``intermediate
asymptotics'' in the language of Barenblatt and
Zel'dovich\cite{barenblatt.1971}.

Since $z$ does not explicitly enter into equation \ref{eqn.12}, it is
convenient to solve the equations by introducing explicitly the mobility flux
$j = -\xi^2 \mbar \partial \mu / \partial z$ and to re-write equation
\ref{eqn.12} as an equation for $j$ as a parametric function of $\mu$ itself,
leaving the $z$ dependence implicit. This gives 

\begin{equation}
  j \partt{j}{\mu} + v j - \bar{\mu} (\mu - \bar{\mu}) \xi^2 \bar{\mu} = 0 
  \label{eqn.13}
\end{equation}

\noindent Outside the mobility generation zone the first two terms involving
mobility transport balance giving 

\begin{equation}
  \begin{split}
  j =& - v \mu + \mathrm{ constant} \\
  j =& - v (\mu - \mu_{\mathrm{low}})
  \end{split}
\end{equation}

\noindent As $\mu$ approaches $\mu_{\mathrm{high}}$, it becomes possible for
the structural rearrangements to take place rapidly enough to lead to mobility
generation. In the mobility generation zone, a different approximation to
equation \ref{eqn.13} is valid in which the first and third terms of equation
\ref{eqn.13} are balanced rather than the first two. The balance in the
generation zone gives an alternate formula for the mobility flux in the
comoving frame: 

\begin{equation}
  \frac{1}{2} \partt{j^2}{\mu} = \bar{\mu}^2 (\mu - \bar{\mu}) \xi^2
\end{equation}

\noindent Integrating this equation at the high mobility side yields

\begin{equation}
  \left[  2 \xi^2 \int_{\mu_{\mathrm{low}}}^{\mu_{\mathrm{high}}} d\mu \bar{\mu}^2 (\mu - \bar{\mu}) \right]^{1/2}
  = j(\mu_{\mathrm{high}})
\end{equation}

\noindent Matching this with the crossover to the mobility transport region
gives an expression for the front velocity

\begin{equation}
  v = \frac{1}{\mu_{\mathrm{high}} - \mu_{\mathrm{low}}} \left\{  
  2 \xi^2 \int_{\mu_{\mathrm{low}}}^{\mu_{\mathrm{high}}} d\mu \bar{\mu}^2(T_F) (\mu - \bar{\mu}(T_F)) 
  \right\}^{1/2}
  \label{eqn.16}
\end{equation}

\noindent To evaluate this expression exactly the fictive temperature profile
needs to be known i.e. how quickly $T_F$ varies in the reaction zone. The $T_F$
profile reflects how far the system is dragged out of local equilibrium through
mobility transport versus local relaxation to the new local thermal
equilibrium. This requires the simultaneous solution of the $T_F$ and $\mu$
equations and must be done numerically. A fairly accurate estimate (an
over-estimate in all likelihood) for the propagation speed can be found,
however.

This estimate for the velocity follows from the fact that the disequilibrium of
mobility $\mu - \mbar(T_F)$ must be smaller than, but can be of the order of,
the overall change in mobility across the front $\mu_{\mathrm{high}} -
\mu_{\mathrm{low}}$.  Introducing this overestimate of the disequilibrium in
the mobility generation zone into equation \ref{eqn.16} gives then

\begin{equation}
  v \approx \frac{1}{(\mu_{\mathrm{high}} - \mu_{\mathrm{low}})^{1/2} }\left\{  
  2 \xi^2 \int_{\mu_{\mathrm{low}}}^{\mu_{\mathrm{high}}} d\mu \bar{\mu}^2(T_F)
  \right\}^{1/2}
  \label{eqn.17}
\end{equation}

\noindent In this expression the behavior near the high mobility end would
dominate. At the high mobility end $\mbar (T_F) \approx \mu$. This relation can
also not be too far wrong at low $\mu$ where again $\mu$ must settle down to
$\mu (T_{\mathrm{low}})$. Using the approximation here that $\mu = \mbar$
uniformly in the integral in equation \ref{eqn.17} yields an estimate valid for
$\mu_{\mathrm{high}} \gg \mu_{\mathrm{low}}$:

\begin{equation}
  v = \sqrt{2/3} \xi \mu_{\mathrm{high}}
  \label{eqn.18}
\end{equation}

\noindent When the overall mobility change is large, the width of the mobility
generating zone $v/\mu$ is thus predicted to be of the order $\xi$, not too
different from the size of an entropic droplet. This scale itself is only
modestly larger than the fundamental molecular size. This suggests, as in the
theory of shock waves where the shock width is of the order of a mean free
path, a more complete microscopic treatment, going beyond the continuum
treatment may be necessary for quantitative accuracy. The microscopic
inhomogeneous BBW-MCT can be solved but it appears to be a daunting numerical
task. Indeed, making the approximation of a molecularly sharp interface would
not be entirely out of the question for analysis. Such a sharp interface would
give a rate of roughly the same magnitude.

We have so far neglected fluctuations entirely in treating front propagation.
Such fluctuations may give rise to instabilities and more complex front
structures. Nevertheless it is the faster reconfiguring regions that should
matter the most. To a first approximation fluctuations should lead to a
velocity which averages the mobility of the rejuvenated sample rather than
averaging the reconfiguration time.

Combining the estimate from equation \ref{eqn.18} with the earlier equation for
the rejuvenation time one obtains

\begin{equation}
  t_R = \left( \frac{\xi_0}{\xi} \sqrt{\frac{3}{2}} \right)^{3/(3+\beta)}
  \left[ \hat{\tau}(T_F^{\mathrm{fin}},T) \right]^{3/(3+\beta)} \left[ \tau (T_F^{\mathrm{in}},
  T) \right]^{\beta / (3+\beta)}
\end{equation}

\noindent Here $\hat{\tau}$ is the harmonic mean relaxation time at the final
fictive temperature $T_F^{\mathrm{fin}} \approx T$.  Since $\xi \approx \xi_0$
the prefactor is of order unity. We thus see the rejuvenation $t_R$ is a
weighted geometric mean of the initial and final relaxation times. We have
assumed in the analysis of the front that the mobile phase has equilibrated
after the passage is complete. It is therefore useful to notice that this
relation confirms that the assumption of stationarity is indeed justified since
the initial relaxation time is indeed much longer than the final one.

More weakly rejuvenated samples (i.e., $T$ greater but close to
$T_F^{\mathrm{in}}$) cannot be analyzed with the simple constant velocity
formula for front propagation. On the other hand a linearized analysis of the
coupled mobility and fictive temperature equations should be adequate to treat
such cases.

\section{Heterogeneous Rejuvenation}

While aging of a macroscopic sample should be dominated by processes in the
bulk, the rejuvenation of a glass should be faster at the surface than it is in
the bulk because the activated processes that originate mobility generation are
faster at the surface. Stevenson and Wolynes have shown that RFOT theory
implies a reduction of the reconfigurational activation free energy by a factor
of two at a free surface in an equilibrated liquid and the same factor should
apply to glasses in the aging regime\cite{stevenson.2008b}.  Mobility
generation will hence proceed much more rapidly at such a free surface and a
rejuvenating front will preferentially start at the surface and propagate into
the bulk. Such heterogeneous rejuvenation was observed by Ediger's group
starting with an ultrastable glass\cite{swallen.2008a}.  They describe the
observed speed of front propagation as having a temperature dependence
paralleling the rate of molecular self diffusion. The present arguments are
consistent with their observations since the self-diffusion constant in a
heterogeneous system averages reconfiguration rates not times, just as the
front velocity here is related to average rates of structural rearrangement.

The stability of front propagation, a central focus of so much combustion
science, may not be too critical for homogeneous rejuvenation since little time
elapses before the fronts merge. The stability analysis is expected to be much
more relevant to the heterogeneous rejuvenation starting from the surface.
Since the effective temperature is not directly transported in the present
model, the situation might seem to resemble combustion of condensed media more
than the combustion of gasses. On this basis, front propagation would be
expected to be stable but clearly a detailed mathematical stability analysis
would be required to establish this, especially when fluctuations in fictive
temperature are explicitly included.

Front propagation from surface to bulk is a well-studied feature of diluent
penetration into glassy polymers, where it is referred to as case II diffusion
\cite{frisch.1980} in which uptake is linear in time, rather than the square
root expected for simple diffusion. A large number of theories of case II
diffusion exist that have many similarities to the present study in purely
mathematical terms\cite{thomas.1982,rossi.1995}. There is a major difference,
however, between those treatments and the present one. Even without a diluent,
that can diffuse, the aging and rejuvenation phenomena studied here by
themselves lead to a diffusion equation for the mobility. If at the same time a
diluent diffuses into the glass, a double diffusion problem appears. If the
diluent molecule is smaller than the molecules in the bulk of the glass, it can
diffuse quite rapidly, thus there will be considerable decoupling of the
diluent diffusion time scale from the glassy $\alpha$ relaxation time which
determines the diffusion of mobility. In combustion the ratio of a reactant's
diffusion coefficient to the kinematic thermal conductivity is known as the
Lewis number\cite{merzhanov.1988,zeldovich.1944,lewis.1934}. The corresponding
ratio of diffusion coefficient to mobility diffusion coefficient $L_{\mu}$ will
be important for type II diluent penetration.

A large Lewis number in combustion leads to unstable, accelerating flame front
propagation often leading to eventual thermal detonation. In analogy, if the
diluent diffusion is strongly decoupled from glassy relaxation i.e. if
$L_{\mu}$ is large, the penetration front will accelerate, leading to more
rapid than linear uptake of the diluent. Although a more complete mathematical
analysis than is done here is needed, the analogy with combustion evident from
the present framework thus provides a natural explanation of accelerating
uptake of penetrating diluents.  Such accelerating uptake has been observed and
is called supercase II diffusion.

\section{Conclusion}

The combination of the generation of mobility by activated reconfiguration
events and mobility transport leads to complex spatiotemporal structures in
glasses. Aging and rejuvenating are seen to be, in no way reciprocal processes
mechanistically\cite{chakrabarti.2004}, although they are described by a common
set of equations. More complex thermal histories can give rise to more complex
spatiotemporal patterns than those described here for simple heating and
cooling experiments. It is hard to believe these patterns are not an essential
aspect of many heat treatment protocols used in technology, but clearly, as in
the study of combustion, the range of phenomena can be quite rich and difficult
to analyze mathematically, even when the best engineering practice is clear
empirically.  Only the simplest cases have been discussed in this paper. In any
event, the present analysis encourages us to contemplate the beautiful shifting
patterns in any piece of glass, flickering like flames, albeit very
majestically.





\begin{acknowledgments}

Inspiring discussions with Mark Ediger, Nate Israeloff, Biman Bagchi and Vas
Lubchenko are gratefully acknowledged. I thank John Deutch also along with them
for critically reading the manuscript. Help preparing the manuscript from Jake
Stevenson and Tracy Hogan is also appreciated.  This work was supported by the
National Science Foundation.

\end{acknowledgments}






\end{article}








\end{document}